\documentclass[graybox]{svmult}
\usepackage{mathptmx}        % selects Times Roman as basic font
\usepackage{helvet}          % selects Helvetica as sans-serif font
\usepackage{courier}         % selects Courier as typewriter font
\usepackage{natbib}

\usepackage{makeidx}         % allows index generation
\usepackage{graphicx}        % standard LaTeX graphics tool
\usepackage{subfig}
\usepackage{multicol}        % used for the two-column index
\usepackage[bottom]{footmisc}% places footnotes at page bottom
\def\ga{\mathrel{\hbox{\rlap{\hbox{\lower4pt\hbox{$\sim$}}}\hbox{$>$}}}}
\def\la{\mathrel{\hbox{\rlap{\hbox{\lower4pt\hbox{$\sim$}}}\hbox{$<$}}}}

\begin{document}
\bibliographystyle{spbasic}
\title*{Substructure and Tidal Streams in the Andromeda Galaxy and its Satellites}
\author{Annette  M. N. Ferguson and A. D. Mackey}
\institute{A. M. N. Ferguson \at Institute for Astronomy, University
  of Edinburgh, Blackford Hill, Edinburgh EH9 3HJ, UK\\
  \email{ferguson@roe.ac.uk} \and A. D. Mackey \at Research School of
  Astronomy \& Astrophysics, Australian National University, Mount
  Stromlo Observatory, Cotter Road, Weston Creek, ACT 2611, Australia\\
  \email{dougal.mackey@anu.edu.au}}
\maketitle

\abstract{Tidal streams from existing and destroyed satellite galaxies
  populate the outer regions of the Andromeda galaxy (M31). This
  inhomogeneous debris can be studied without many of the obstacles
  that plague Milky Way research.  We review the history of tidal
  stream research in M31, and in its main satellite galaxies.  We
  highlight the numerous tidal streams observed around M31, some of
  which reside at projected distances of up to $\sim 120$~kpc from the
  center of this galaxy.  Most notable is the Giant Stellar Stream, a
  signature of the most recent significant accretion event in the M31
  system.  This event involved an early-type progenitor of mass $\sim
  10^9$~M$_{\odot}$ that came within a few kpc of M31's center roughly
  a gigayear ago; almost all of the inner halo (R~$\leq 50$~kpc)
  debris in M31 can be tied either directly or indirectly to this
  event.  We draw attention to the fact that most of M31's outer halo
  globular clusters lie preferentially on tidal streams and discuss
  the potential this offers to use these systems as probes of the
  accretion history.  Tidal features observed around M33, M32, NGC~205
  and NGC~147 are also reviewed.  We conclude by discussing future
  prospects for this field.}

\section{Introduction}
\label{sec:intro}
 
Within the context of the cold dark matter paradigm, structure
formation proceeds hierarchically and galaxies like the Milky Way and
Andromeda (M31) are predicted to arise from the merger and
accretion of many smaller sub-systems as well as from the smooth
accretion of intergalactic gas \citep[e.g][]{whi78, whi91}.  Galaxy
outskirts are of particular interest since the long dynamical
timescales in these regions mean that coherent debris from past
accretion events has the greatest longevity.  The discovery of the
Sagittarius dwarf galaxy and associated tidal stream (see Law \& Majewski,
this volume)
demonstrated beyond a doubt that satellite accretion played an
important role in the growth of the Milky Way's halo, but the veracity
of this aspect of the hierarchical model could not rest solely on a
single event observed within a single galaxy. In this spirit, the late
90s saw the quest begin to identify other galactic systems which were
visibly in the process of devouring smaller satellites.

Our nearest giant neighbour, M31, provides the most obvious target for
such studies.  Lying at $\sim780$~kpc, it is in many respects a sister
galaxy to the Milky Way.  It has a similar total mass
\citep[e.g.][]{dias14,vel14}, is of a similar morphological type, and
it resides in the same low-density environment -- one which is deemed
typical for much of the present-day galaxy population.  M31's
proximity means that individual stars near the tip of the red giant
branch (RGB) can be resolved from the ground; this offers a
powerful method for probing very low effective surface brightnesses,
such as those expected for tidal debris streams.  Furthermore, its
high inclination to the line-of-sight ($i \sim 77^{\circ}$) means that
it is ideally suited for studies of its halo regions.

There are some disadvantages to studying tidal streams in M31 as
opposed to our own Milky Way.  Even with the world's largest
telescopes, ground-based studies of M31's halo are limited to using
luminous giant stars whereas Milky Way studies can harness the power
of the much more numerous main sequence turn-off population
\citep[e.g.][]{bel06}.  Additionally, in M31 we are mainly confined to
analysis of projected positions on the sky (although occasionally some
line-of-sight distance information is available), and we can only
measure radial velocities. This can be contrasted with the situation
in the Milky Way where it is possible to additionally determine
line-of-sight distances\footnote{In practice, individual distances to
  large samples of stars in the Milky Way are crude at the moment.
  ESA's {\it Gaia} mission will change this when it starts to deliver
  data in 2016 but the most accurate distances will be limited to
  stars within roughly 10~kpc of the Sun.} and proper motions,
and thus probe the full six dimensional phase space.  In effect, these
differences imply that in M31 we are sensitive to tidal streams that
are of higher surface brightness than those we can uncover in the
Milky Way, and moreover typically only the subset retaining a high
degree of spatial and/or kinematical coherence.

On the other hand, there are some clear advantages to studying an
external system such as M31. Our vantage point largely alleviates
complicated line-of-sight projection and extinction effects, such as must be endured
in studies of the Milky Way.  This means that we have a better
understanding of the morphology of tidal features and where
stellar substructures lie (at least in projection) with respect to
each other, and with respect to additional halo tracers such as
globular clusters (GCs) and dwarf satellites. Our bird's-eye view also
makes it fairly straightforward to construct {\it in situ} samples of
halo stars at various radii.  Remarkably, we currently know more about
the outer halo (R~$\ga50$~kpc) of M31 than we do of the Milky Way. In
the Milky Way, the outer halo is obscured by a dense veil of
foreground stars making the robust identification of the low density
population of outer halo stars difficult.  In addition, while the main
sequence turn-off method so extensively used by the Sloan Digital Sky
Survey (SDSS) teams has probed out to distances of $\sim40$~kpc, such
stars are not detected in SDSS imaging at distances beyond this.
While some outer halo substructures have been uncovered using other
tracers \citep[e.g.][]{new03,wat09}, the inhomogeneous and sparse
nature of these studies precludes any meaningful conclusions about the
global properties of the Milky Way's outer halo.  In M31, we also have
a far clearer view of the low latitude regions of the galaxy, enabling
discrimination between perturbed disk features and accreted
substructures.
 
Over the last 15 years, a multitude of studies have targeted the
outskirts of M31; this chapter reviews the tremendous progress and exciting results that have emerged from this work.

\section{Historical Studies}
\label{sec: hist}

While the first detailed studies of faint structure and substructure
in M31 did not appear until the early 2000s, there were a number of
studies prior to this which are of particular note.

\citet{baa63} were the first to comment on the strong warp present in
the outer parts of M31's stellar disk.  Baade (p. 73) notes that the
disk signature is still present at a major axis distance of
$2^{\circ}$ but that the opposing sides of the disk have $``$swirled"
off in anti-symmetric directions by a radius of 2.25$^{\circ}$ -- an
observation he speculates could be due to the tidal action of the
Milky Way on M31.  The existence of the prominent stellar warp was
further confirmed by \citet{inn82} who stacked digital Palomar Schmidt
plates to reach surface brightnesses in M31 of $\mu_{\rm V} \ga 25.8$
mag~arcsec$^{-2}$.  Early H$\,${\sc i} studies of M31 also revealed a
strong warp in the neutral hydrogen disk
\citep[e.g.][]{rob75,new77,cram80}.  While the stellar and H$\,${\sc
  i} disks are warped in the same direction, the stellar warp appears
to begin at a smaller radius than the H$\,${\sc i} warp and it
exhibits a greater deviation from the disk plane; this holds true
along both major axes but seems especially apparent in the
north-east.

\citet{wk88} conducted the definitive study of the light distribution
in M31 before the era of wide-field resolved star mapping, including a
first quantitative exploration of the peripheral disk.  They
constructed multi-band images from digitised Burrell Schmidt plates
and analysed surface brightness and color profiles across the disk.
They also detected a clear warp, but speculated that the north-eastern
warp, due to its faintness and extreme bend, may actually be a
galactic reflection nebula and not a stellar feature associated with
M31. This issue was settled by \citet{mor94} who resolved the stellar
populations in the north-eastern warp (which they termed the `Spur')
for the first time and showed they lay at the distance of M31.  Fig. 1
of \citet{wk88} shows beyond doubt that the north-eastern half of the
disk is far more perturbed than the south-western half, and, in
hindsight, one can even see a slight luminosity enhancement in the
direction of the Giant Stellar Stream.
 
Equally influential were the first studies of M31's resolved stellar
halo.  \citet{mk86}, \citet{cro86} and \citet{pvdb88} all presented
colour-magnitude diagrams (CMDs) of small regions located at several
kpc along M31's minor axis, a region perceived at the time to be
dominated by pure halo.  These studies revealed a moderate metallicity
population ([Fe/H]~$\sim -1$) with a large metallicity spread, in
stark contrast to the metal-poor halo of the Milky Way.  The disparity
of the halo populations in two galaxies that were otherwise considered
rather similar remained a significant puzzle throughout the following
decade.

\section{Wide-Field Mapping Surveys of M31}
\label{sec: wfsurveys}

The relative proximity of M31, while advantageous for detailed study,
also poses a problem in that even the main body of the galaxy subtends
a substantial angle on the sky. In the mid-to-late 90s, wide-format
CCD detectors became increasingly available on medium-sized telescopes
and this development opened up new possibilities for surveying the
faint outlying regions of M31.  With a distance modulus of
$m-M=24.47\pm 0.07$ \citep{mcc05}, stars near the tip of M31's RGB
($M_I\approx-4$) have $I\approx20.5$ and thus could be easily detected
in modest exposures with a 2-m class telescope.  As a result, it
became feasible to contiguously map resolved stars at the bright end
of the luminosity function over very large areas.

RGB stars are the evolved counterparts of low to intermediate mass
(M~$\sim0.3-8$ M$_{\odot}$) main-sequence stars with ages of at least
$\ga 1$ Gyr.  While asymptotic giant branch stars and high mass main
sequence stars can be even more luminous than RGB stars, RGB stars are the
most interesting in the context of tidal stream research since they
are long-lived and can be used to trace the old stellar components of
galaxies, where signatures of hierarchical assembly are expected to be
most prevalent.  Additionally, for a fixed age, the colour of an RGB
star depends almost entirely on its metallicity; thus, for a roughly
uniform age for a stellar population, individual stellar metallicities
can be derived purely from photometry.

\begin{figure}[h]
\begin{center}
\subfloat{{\includegraphics[height=5.5cm]{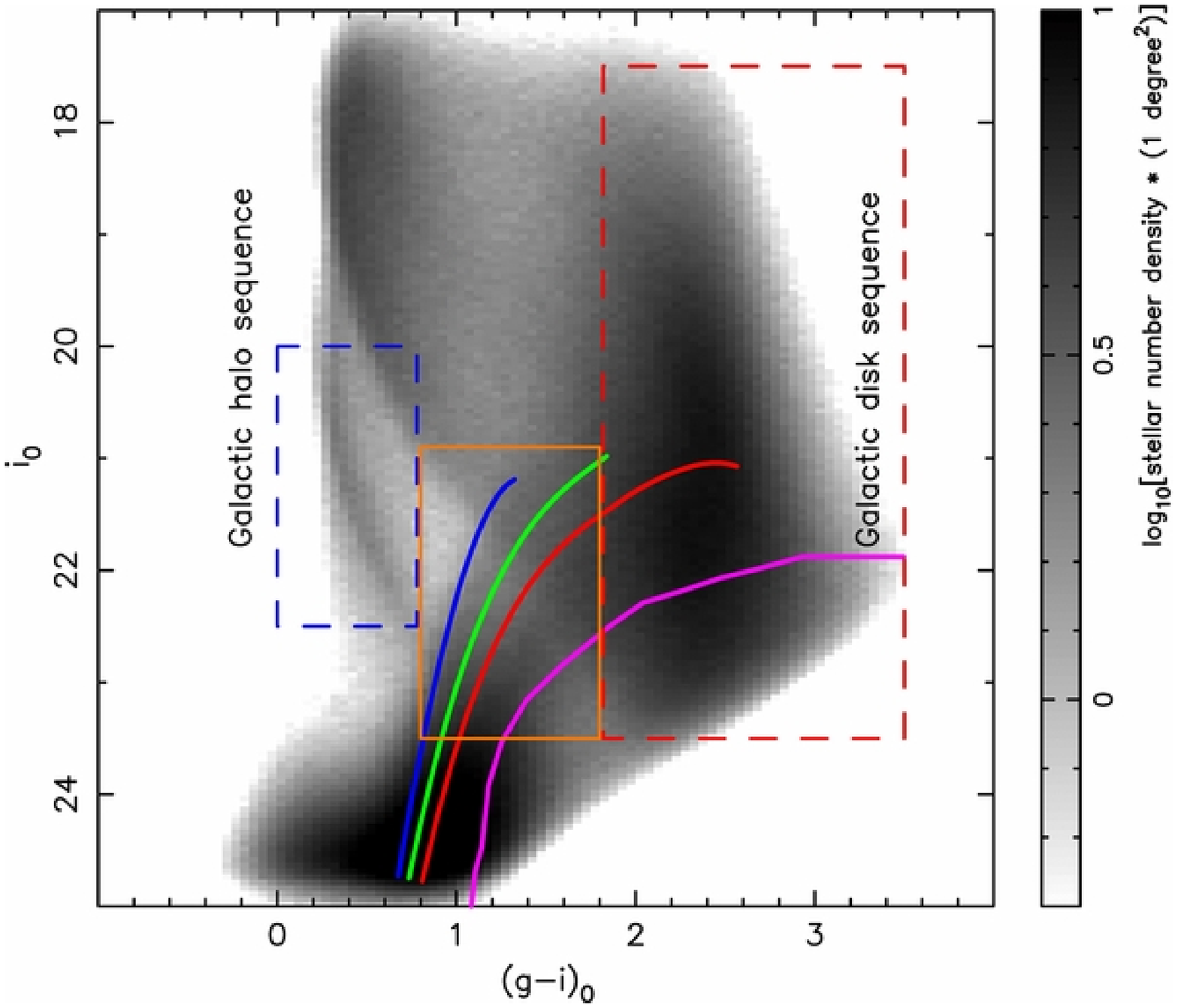}}}
\subfloat{{\includegraphics[height=5.5cm]{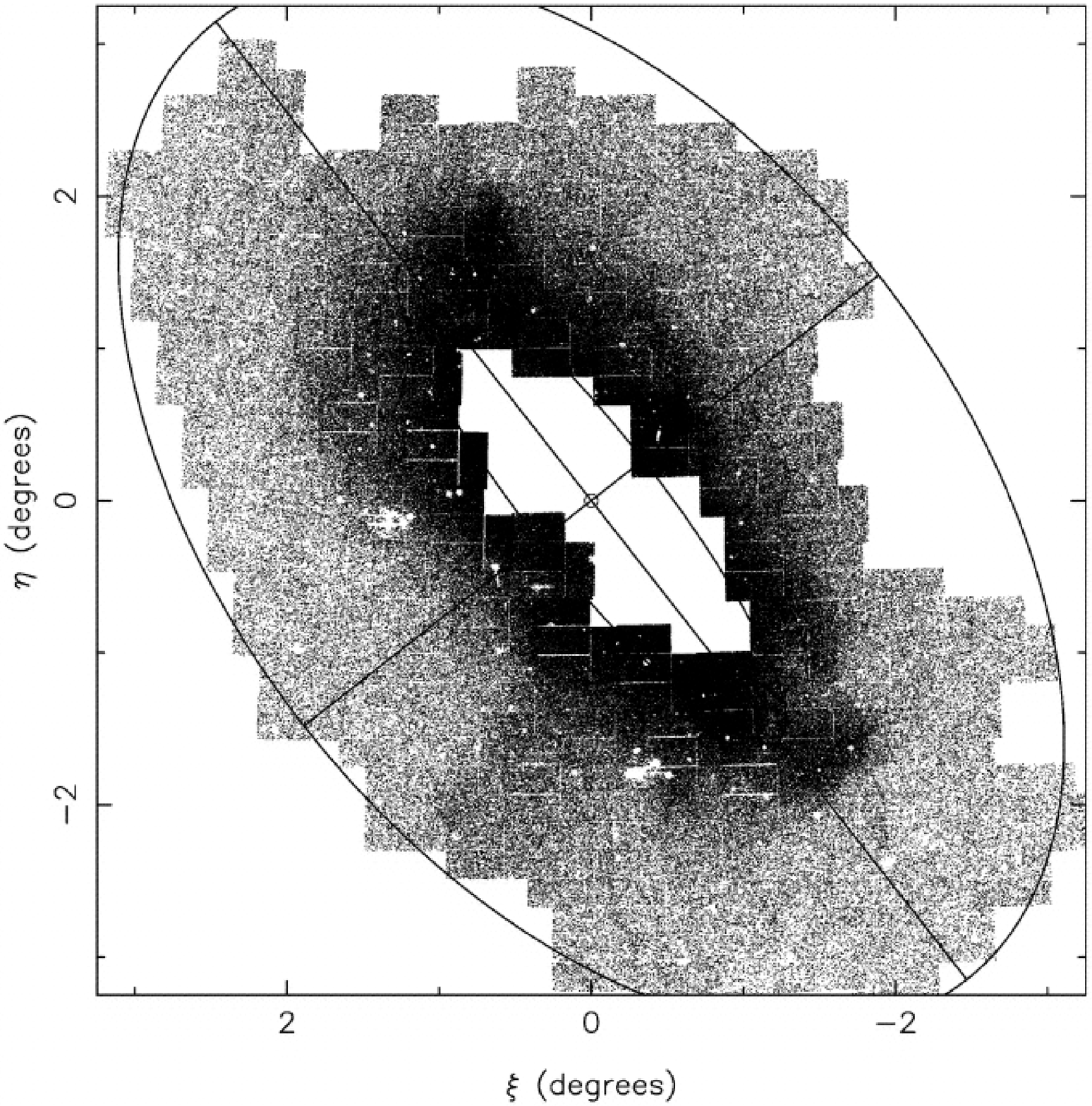}}}
\end{center}
\caption{(Left) A Hess diagram of the point sources in the PAndAS
  survey at distances beyond $2^{\circ}$ of M31 \citep[reproduced 
  from][]{iba14} .  A series of fiducial tracks spanning [Fe/H]$=
  -1.91, -1.29, -0.71$ and $-0.2$ are superimposed on the RGB while
  contaminating Milky Way foreground disk and halo sequences are
  indicated by dashed boxes.  Blueward of $ (g-i)_0\sim 1$ and fainter
  than $i_0\sim23$, unresolved background galaxies become the primary
  contaminant.  The orange box shows the adopted colour-magnitude
  selection for M31 RGB stars.  (Right) An early map of metal-poor RGB
  star counts around M31 from the INT/WFC survey \citep[reproduced
  from][]{fer02}. The outer ellipse has a flattening of 0.6 and a
  semimajor axis length of 55~kpc.}
\label{fig:hess+int}
\end{figure}

Resolved star surveys map the spatial distribution of individual RGB
stars which can in turn be used to infer the surface brightness
distribution of the underlying light.  This technique allows much
lower surface brightness levels to be reached than typically
achievable with conventional analyses of diffuse light.  As a
simplistic illustration of how the method works, consider a population
of M31 RGB tip (TRGB) stars with $I_0 \sim20.5$ and $(V-I)_0 \sim
1.5$.  A surface density of $10^5$ such stars per square degree
corresponds to $\mu_{V} \approx 27$ mag~arcsec$^{-2}$ while a surface
density of $10^3$ stars per square degree corresponds to $\mu_{V}
\approx 32$ mag~arcsec$^{-2}$.  This calculation is crude since it
neglects the fact that there is a range of RGB luminosities within a
population, and also that some sizeable fraction of the total light
will come from stars fainter than the magnitude limit, but these are
corrections that can be easily calculated for any given survey
\citep[e.g.][]{pvdb94,mcc10}.  Nonetheless, it is sufficient to
demonstrate that, in the low crowding outer regions of M31 (and
external galaxies in general), the resolved RGB star technique is
clearly the optimal means to search for and map very faint structures.
The main challenges in using this method are to image to sufficient
depth to detect a statistically significant population of RGB halo
stars, and to disentangle genuine RGB stars from contaminating
populations, which typically consist of foreground Milky Way dwarf
stars and unresolved background galaxies (see the left panel of Fig.
\ref{fig:hess+int}).

The breakthrough in our ability to search for low surface brightness
structure in the outskirts of M31 led to the discovery of a plethora
of faint streams and substructure in the the inner halo, including the
dominant Giant Stellar Stream.  The pioneering INT Wide Field Camera
survey \citep[e.g.][]{iba01,fer02, irw05} mapped $\approx 40$ square
degrees (163 contiguous pointings) around M31 in the $V$ and $i$
passbands, reaching to $\sim 3$ magnitudes below the TRGB (see 
the right panel of Fig.
\ref{fig:hess+int}).  M31 was also targeted by the SDSS
\citep{zuc04a}, where stars near the TRGB were mapped to large
distances along the major axis.  These surveys also uncovered several
previously-unknown M31 dwarf satellites and globular clusters
\citep[e.g.][]{zuc04b,zuc07,irw08, hux08}.

In parallel to these efforts to survey RGB stars, other groups began
to explore the halo and outer disk populations using planetary nebulae
(PNe) \citep[e.g.][]{mer03, mer06, mor03, kni14}.  Although PNe offer
some advantages over RGB stars as tracers of halo light (for example,
they are more luminous, suffer much reduced sample contamination, and
can provide simultaneous information on radial velocities), they are
far rarer.  From a survey of PNe in the M31 bulge, \citet{cia89} found
that the ratio between the number of PNe in the top 2.5 magnitudes of
the luminosity function and the $V$-band luminosity was
$\alpha_{2.5}\sim 30.8 \times 10^{-9}$ PNe per L$_{\odot}$.  This
leads to the expectation of $\approx50$ PNe per square degree at
$\mu_{V}\sim 24$ mag~arcsec$^{-2}$ but only $\approx1$ per square
degree at $\mu_{V}\sim 28$ mag~arcsec$^{-2}$.  This makes surveys for
resolved RGB stars far more efficient than those for PNe in the outer
regions of M31.  However, the situation changes for more distant
galaxies, where RGB stars become too faint to resolve while PNe can
still be detected \citep[e.g.][]{coc13,fos14}.

Based on the success of these early studies, exploration of the M31
{\it outer} halo began in the mid 2000s. Due to the lower stellar
density in these parts, deeper photometry was required in order to
sufficiently sample the RGB luminosity function and this in turn
required larger telescopes.  The Pan-Andromeda Archaeological
Survey (PAndAS) was conducted using the MegaCam instrument on the
3.8-m CFHT to contiguously map over 380 square degrees around the
M31-M33 region and detect stars to $\sim4$ mag below the TRGB
\citep[e.g.][]{iba07, mcc09, iba14}.  Other wide-field ground-based
work concentrated on deep pencil beam studies of the outer halo
\citep[e.g.][]{ost03, tan10}.

\section{Major Tidal Features in the Halo of M31}
\label{sec:substructure}

\begin{figure}[h]
\begin{center}
\includegraphics[width=8cm]{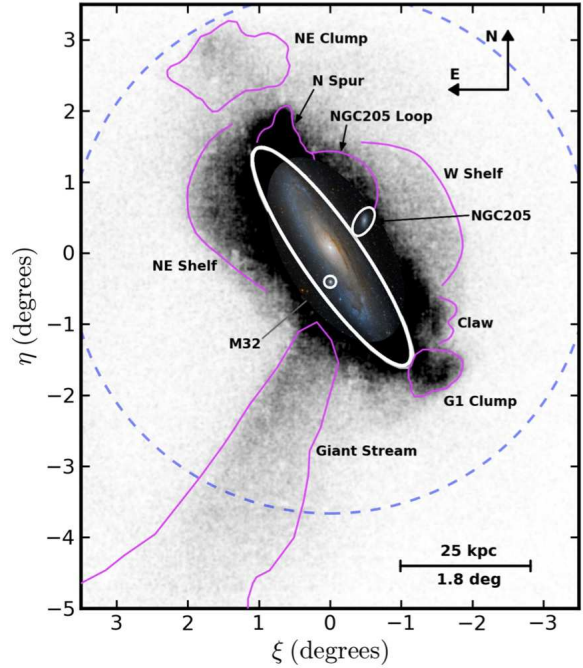}
\end{center}
\caption{The PAndAS map of metal-rich RGB stars in the inner halo of
  M31, upon which a typical textbook image of M31 is superposed. The
  map is constructed from stars with $i_0\le 23.5$, having
  $-1\le$~[Fe/H]~$\le 0$.  The large white ellipse has a semi-major axis of
  27~kpc and delineates the full extent of the bright disk; the dashed
  blue circle has a radius of 50~kpc.  Prominent inner halo
  substructure is outlined and labelled, as are the dwarf satellites
  M32 and NGC~205.  }
\label{fig:inner_map}
\end{figure}

\begin{figure}[h]
\begin{center}
\includegraphics[width=\textwidth]{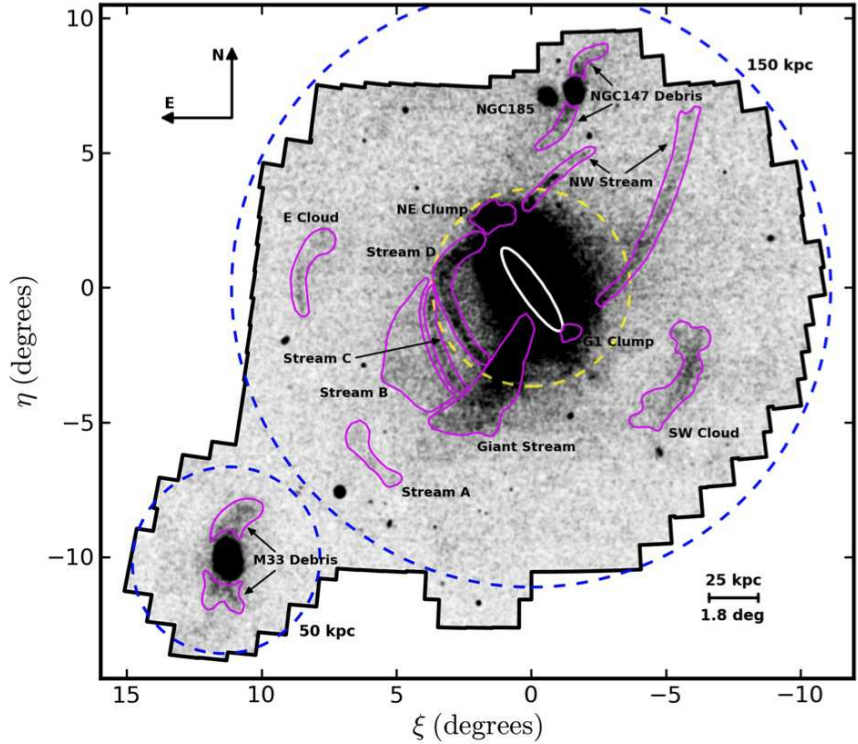}
\end{center}
\caption{A map of metal-poor RGB stars across the full extent of the
  PAndAS survey ($i_0\le 23.5$ and $-2.5\le$~[Fe/H]~$\le-1.1$). The
  white ellipse is the same as in Fig. \ref{fig:inner_map} while the
  dashed circles surrounding M31 have radii of 50~kpc (yellow) and
  150~kpc (blue) respectively and that surrounding M33 has a radius of
  50~kpc.  Prominent outer halo substructure is outlined and labelled,
  as are tidal streams associated with the satellite galaxies M33 and
  NGC~147.}
\label{fig:outer_map}
\end{figure}

\begin{figure}[h]
\begin{center}
\includegraphics[width=\textwidth]{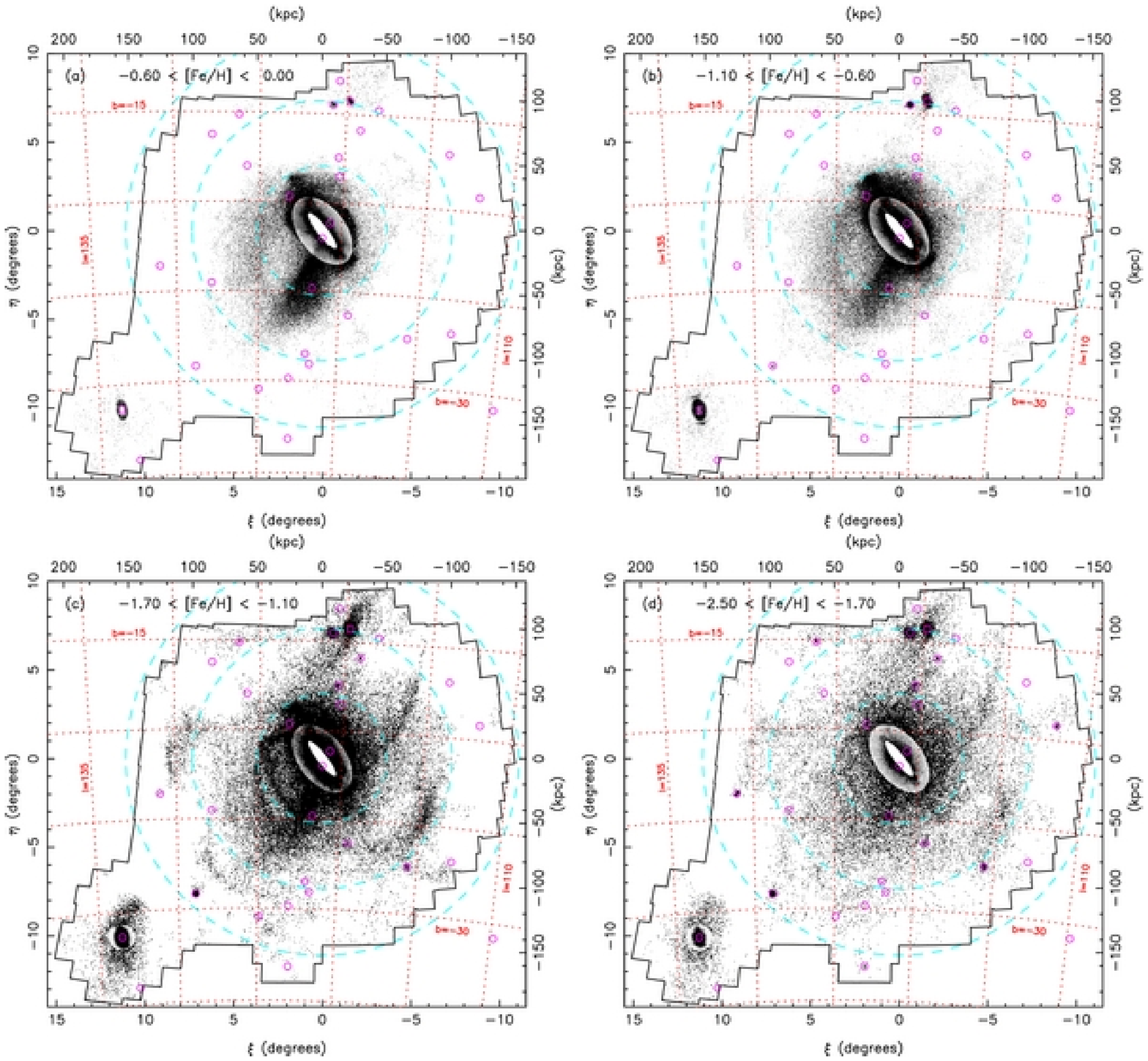}
\end{center}
\caption{The PAndAS map of RGB stars, as in Fig.~\ref{fig:outer_map},
  but presented this time as a function of metallicity.  The two high
  metallicity bins (top panels) are dominated by the Giant Stellar
  Stream, although note that this structure changes morphology
  slightly between the two panels. While the low metallicity bottom
  left panel is dominated by numerous streams, the more metal-poor
  right-hand panel appears much smoother. Reproduced from \citet{iba14}.}
\label{fig:rod_halo}
\end{figure}

Figs. \ref{fig:inner_map} and \ref{fig:outer_map} show maps from the
PAndAS survey of RGB stars in the inner and outer halo regions of M31,
displayed to highlight prominent substructures.  Additionally, Fig.
\ref{fig:rod_halo} shows the distribution of RGB stars in four
different metallicity bins, revealing how the morphology of the tidal
debris changes as a function of metallicity.  A striking feature of
all of these visualisations is the non-uniformity of the stellar
distribution in the outer regions. The most metal-poor map presented in Fig.
\ref{fig:rod_halo} has a smoother appearance than the others, but it
still exhibits a substantial degree of substructure.

The inner halo (R~$< 50$~kpc) of M31 appears as a 
flattened structure (axis ratio $\sim 0.5$) in Fig.
\ref{fig:inner_map}, around the edge of which bright tidal features
(e.g. streams, clumps, spurs, shelves) can be seen. In the outer halo
(R~$\ge 50$~kpc), the most prominent features are a multitude of faint
narrow streams and arcs. Based on their appearance in Fig.
\ref{fig:rod_halo}, these outer streams are also considerably more
metal-poor than the substructures which dominate the inner halo.
\citet{lew13} searched for a correlation between the tidal structures
seen in stars, and features in the H$\,${\sc i} gas around M31.
Interestingly, they found a general lack of spatial correlation
between these two components on all scales, with very few potential
overlaps.

A brief description of some of the most prominent tidal features seen
around M31 is given below:
 
 \begin{itemize}
 \item {\bf Giant Stellar Stream (GSS):~~} Discovered in the
   first quadrant of M31 that was mapped by the INT/WFC survey, the
   GSS is the most prominent overdensity in M31's halo and covers a
   large fraction of its south-east quadrant \citep{iba01}.  It
   can be traced as a coherent structure to a projected galactocentric
   radius of $\sim 100$~kpc, and spans a width of $\sim 25$~kpc.  The
   stream has a linear morphology with a sharp eastern edge and an
   estimated absolute $V$-band magnitude of $M_V \approx -14$
   \citep{iba01}. However, this is a crude estimate based on only that
   part of the stream which is visible in the earliest INT maps.  As
   the stream is now known to be more than twice as long as this, and
   other debris features have been identified as forward wraps of the
   structure, the total luminosity of the GSS could easily be 1--2
   magnitudes higher.  \citet{iba07} show that there is a large-scale
   stellar population gradient present, with the high surface
   brightness core region of the stream having relatively more
   metal-rich stars than the peripheral regions. Both photometric and
   spectroscopic studies reveal the core stream to have a moderately
   high metallicity of [Fe/H]~$\geq-0.5$ to $-0.7$, with the envelope
   dropping to [Fe/H]~$\sim-1.4$ \citep{guh06,iba07,gil09}.
 
 \item{\bf G1 Clump:~~} This feature was first recognised in the
   INT map published by \citet{fer02} and appears as a rather round
   clump of stars located at a projected radius of $\sim 30$~kpc along
   the south-western major axis of M31.  It has dimensions of
   $0.5^{\circ} \times 0.7^{\circ}$, or $7 \times 10$~kpc at the
   distance of M31.  \citet{fer02} estimate an absolute magnitude of
   $M_V \approx -12.6$ and a $V$-band surface brightness of $\approx
   28.5$ mag~arcsec$^{-2}$. The feature was originally named because
   the luminous M31 globular cluster G1 lies nearby. This star
   cluster is notable because it has been argued to have both an
   internal metallicity spread as well as an intermediate-mass black
   hole, characteristics that suggest it could be the remnant core of
   a nucleated dwarf elliptical galaxy \citep[e.g.][]{mey01,geb05}.
   While the detection of tidal debris in the vicinity of this
   enigmatic object was very exciting, subsequent observations of the
   properties of stars in the G1 Clump appear to rule out any
   association between the two \citep{rich04,rei04,iba05,faria07}.
 
 \item{\bf North-East Clump (NE Clump):~~} Located at a
   projected radius of $\sim 40$~kpc and near the north-eastern major
   axis, this substructure is one of the most nebulous features in the
   inner halo of M31.  It subtends a diameter of $\sim 1^{\circ}$ or
   $\sim 14$~kpc at the distance of M31, and appears to connect to the
   main body of the galaxy by a faint filament.  It is estimated to
   have an absolute $g$-band magnitude of $M_g \approx -11.6$ and a
   $g$-band surface brightness of $\approx 29.0$ mag~arcsec$^{-2}$
   \citep{zuc04a}.  Although it was initially suggested that the NE
   Clump was a disrupting dwarf satellite, subsequent observations
   have disfavoured this interpretation \citep{iba05,ric08,ber15a},
   since the stellar populations are more representative of the disk.
   
 \item{\bf North-Eastern (NE) and Western (W) Shelves:~~} The NE
   Shelf is a diffuse but fairly sharp-edged extension lying
   north-east of M31's center, while the Western Shelf is a
   fainter feature of similar morphology and size on the opposite side
   of the galaxy.  On the basis of their comparably high
   metallicities, \citet{fer02} suggested the NE Shelf could be an
   extension of the GSS.  Using inferences from $N$-body simulations,
   \citet{far07} argued that both the NE and W Shelves were forward
   continuations of the stream, representing material stripped off
   during successive pericentric passages.
 
 \item{\bf Streams B, C, D:~~} Identified by \citet{iba07}, Streams
   B--D are a series of approximately parallel tangential streams
   which cross the southern minor axis of M31 inside a radius of 100
   kpc. Their metallicities are in the range $-1.5 \leq \rm{[Fe/H]}
   \leq -0.5$.  The eastern portions of Streams C and D appear to
   overlap in projection and all of the streams seem to terminate, or
   at least dramatically fade, once they reach the GSS.  Stream
   C has been shown to be particularly complex, consisting of two
   distinct (but overlapping) metallicity and kinematic components
   \citep{chap08, gil09}.  Along with the GSS, the metal-rich
   component of Stream C is the only other outer halo feature visible
   in the most metal-rich maps of Fig. \ref{fig:rod_halo}.
  
 \item{\bf Far Outer Halo Streams (R~$\ga 100$~kpc):~~} Figs.
   \ref{fig:outer_map} and \ref{fig:rod_halo} show that the outer halo
   is littered with various faint streams and clumps, including
   Stream A, the North-West (NW) Stream, the South-West
   (SW) Cloud))) and the (((Eastern (E) Cloud \citep{iba07, mcc09,
     ric11, car11, bate14}.  With $\mu_{\rm V} \geq 31.5$
   mag~arcsec$^{-2}$, these streams represent the faintest
   spatially-coherent debris yet identified around M31.  Stream A
   crosses the minor axis at a projected radius of $\sim 120$~kpc and
   aligns with the inner network of tangential streams.  The E and SW
   Clouds appear as stellar arcs at slightly smaller radii on either
   side of Stream A.  The NW Stream is a long ($\sim100$~kpc) and
   narrow ($\sim3$~kpc) radial feature; although there is no visible
   connection between the upper and lower branches of the stream, the
   fact that both trace out segments of a single ellipse and have a
   similar metallicity has supported the notion they are related
   \citep{car11}.  Although part of the NW Stream was also seen in the
   deep pencil beam study of \citet{tan10} (their (Stream F), their
   Stream E was not recovered in the PAndAS survey.

\end{itemize}

\section{Understanding the Nature and Origin of Tidal Features in M31}

Following the initial discovery of tidal features in the M31 halo,
concerted follow-up observations and detailed modelling have been
carried out in order to develop a complete understanding of this
material. In this section, we review and summarise some highlights
from this work.
 
\subsection{The Giant Stellar Stream}
\label{sec:gss}
Over the last decade, the GSS has been the subject of intense study.
On the observational side, efforts have concentrated on deriving
quantities (e.g. distance, velocity) that can be used to model the
orbit of the progenitor and on stellar populations constraints that
can be used to establish its nature. On the theoretical side, work has
focused on reconstructing the orbital history of the progenitor and
using this knowledge to measure the halo mass of M31.

Line-of-sight distances to the stream are one of the key inputs for
GSS orbit models. \citet{mcc03} used measurements of the TRGB in a
series of CFHT12K pointings to show that the GSS lies $\ga100$~kpc
behind M31 at a projected radial distance of 60~kpc and moves
progressively closer to galaxy at smaller radii, with their distances
being indistinguishable (within the uncertainties) at $\leq 10$~kpc.
These authors were also able to detect the stream in two fields on the
northern side of M31, where it lies $\sim 40$~kpc in front of the
galaxy, but no further, suggesting that the stream wraps fairly
tightly around the M31 center towards the north-east.

Another key observable is the radial velocity distribution of stream
stars.  Individual RGB stars at the distance of M31 are sufficiently
faint that an 8-m class telescope is required in order to measure
their line-of-sight velocities; the DEIMOS multi-object spectrograph
on the Keck \rm{II} telescope has been the source of almost all the GSS
radial velocity measurements to date \citep{iba04,guh06, kal06,gil09}.
These studies have shown that the radial velocity of the GSS becomes
increasingly positive with increasing distance from the center of M31,
ranging from $v_{helio}=-320$~km~s$^{-1}$ at a projected distance of
60~kpc to $v_{helio}=-524$~km~s$^{-1}$ at a projected distance of
17~kpc.  In all fields studied so far, the velocity dispersion is
quite narrow -- in the range of $10-30$~km~s$^{-1}$.  Intriguingly, a
second cold kinematic component has been detected at several locations
along the GSS \citep{kal06,gil09}. It has the same velocity gradient
(and dispersion) as the primary GSS over the range in which the two
have been mapped ($\sim 7$~kpc) but it has a radial velocity that is
offset by $\sim+100$~km~s$^{-1}$. It is presently unclear whether this
component is due to M31's disturbed disk or a forward wrap or
bifurcation of the main stream.

The combination of line-of-sight distance and radial velocity
measurements indicates that the GSS progenitor fell almost straight
into M31 from behind.  These data have motivated a variety of efforts
to model the accretion of a dwarf satellite galaxy on a highly radial
orbit \citep{iba04, font06, far06, far08, far13, mori08, sad14}.
While these models differ in various aspects, they generally agree on
the fact that the progenitor's initial stellar mass was in the range
$1-5 \times 10^9$~M$_{\odot}$ and that its first pericentric passage
came within a few kpc of the M31 center less than 1-2~Gyr ago.  Some
properties of the observed GSS, in particular the asymmetric
distribution of stars along the stream cross section and the internal
population gradient, are better reproduced in $N$-body models in which
the progenitor possessed a rotating disk \citep[e.g.][]{far08, sad14}.

\begin{figure}[h]
\begin{center}
\includegraphics[width=\textwidth]{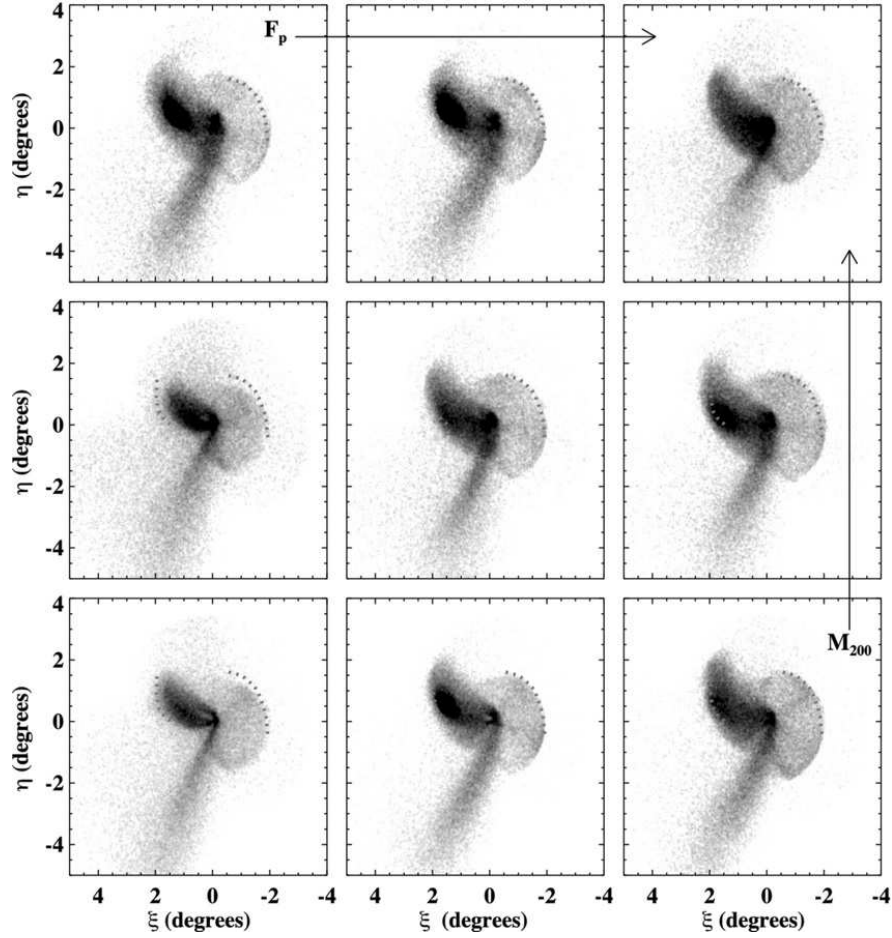}
\end{center}
\caption{Spatial distribution of stellar debris from the tidal
  disruption of the GSS progenitor, as predicted by a set of 9
  $N$-body simulations.  The dashed lines show the observed boundaries
  of the NE and W shelves.  In each case, different combinations of
  $M_{200}$, the virial mass of M31, and $F_{\rm{p}}$, the orbital
  phase of the progenitor at the present epoch, are adopted.  While
  all panels share similar morphologies and resemble the main features
  seen in Figs. \ref{fig:inner_map} and \ref{fig:outer_map}, some
  systematic differences can be seen as a function of the parameters.
  The present-day state of the progenitor varies from being tightly
  bound to highly dispersed; in all models shown here, it lies in the
  region of the NE shelf. Reproduced from \citet{far13}.}
\label{fig:fardal_sims}
\end{figure}

In this near head-on collision, the progenitor experiences significant
destruction at the first pericentric passage.  Much of the satellite's
mass is stripped off to form leading and trailing tidal streams, and a
generic prediction is that much of M31's inner halo should be littered
with this debris.  Fig. \ref{fig:fardal_sims} shows the sky
distribution of debris in a set of recent $N$-body models of the GSS's
orbit within the M31 potential \citep{far13}.  Each panel differs in
the adopted values of $M_{200}$, the M31 mass inside a sphere
containing an average density 200 times the closure density of the
universe, and $F_p$, the orbital phase of the progenitor at the
present day.  Although the exact pattern of debris depends on these
(and a few other) parameters, all panels exhibit a similar morphology
and show remarkable consistency with some of the features seen in
Figs.  \ref{fig:inner_map} and \ref{fig:outer_map}.  
The GSS, the NE Shelf and the W Shelf are all naturally reproduced in
this scenario, with the GSS representing the trailing stream of
material torn off in the progenitor's first pericentric passage while
the NE and W shelf regions contain material torn off in the second and
third passages, respectively.  These predictions are in excellent
agreement with observations of the stellar populations of these
substructures  -- which show striking
similarities in their star formation histories (SFHs) and metallicity distributions 
to those of the GSS \citep[e.g.][]{fer05,ric08} -- as well as with the observed 
kinematics of the shelves \citep[e.g.][]{gil07,far12}.  The extensive
pollution of M31's inner halo by material stripped from the GSS
progenitor provides at least a partial explanation for why this region
is dominated by more metal-rich stellar populations compared to the
Milky Way \citep[e.g.][]{mk86, iba14}.

Based on the projected positional alignment with the GSS, M31's
luminous dwarf elliptical satellites M32 and NGC~205 were
initially considered as prime candidates for the progenitor
\citep{iba01,fer02}.  Indeed, both these systems possess unusual
properties and, as will be discussed in Section \ref{sec:sats}, are
tidally distorted in their outer regions \citep{choi02,fer02}.
However, even the earliest attempts at orbit models ruled out a
straightforward connection between the GSS progenitor and either of
these two satellites \citep{iba04}\footnote{Meanwhile, \citet{blo06}
  argue that a head-on collision between M31 and M32 about 200 million
  years ago could be responsible for the formation of two off-center
  rings of ongoing star formation seen in the M31 disk.}.  Current
models agree that any existing remnant should lie in the region of the
NE Shelf, although thus far no candidate has been identified.

Although the location of the GSS progenitor remains a mystery,
analysis of the stellar populations in the stream has provided
important insight into its nature.  Fig. \ref{fig:jenny_cmds} shows
deep HST CMDs of a variety of tidal debris fields in the inner halo of
M31, including those associated with the GSS.  One can clearly see two
predominant CMD morphologies -- those which contain a narrow tilted
red clump and a prominent horizontal branch extending quite far to the
blue (labelled $`$SL' in Fig. \ref{fig:jenny_cmds}), and those which
contain a rounder red clump, a well-populated blue plume and exhibit
no horizontal branch (labelled $`$DL' in Fig. \ref{fig:jenny_cmds}).
Those pointings which directly sample material associated with the GSS
progenitor uniformly exhibit the morphology of the former type
\citep{fer05,ric08}.  A third category, labelled $`$C', appears to be
a composite of the previous morphologies.

Quantitative measures of the SFH and age-metallicity relation (AMR) of
the GSS have been derived from deep CMDs using synthetic modelling
techniques \citep{bro06,ber15a}.  The bottom panels of Fig.
\ref{fig:sfhs} show the most extensive analysis carried out to date,
based on the 5 fields lying on GSS debris located throughout the inner
halo.  It appears that star formation in the progenitor got underway
early on and at a fairly vigorous pace, peaking 8--9 Gyr ago.  Star
formation remained active until about 6~Gyr ago when there was a very
rapid decline; this $`$quenching' may indicate the time when the
progenitor first entered the halo of M31.  Roughly 50\% of the stellar
mass in the GSS fields was in place $\sim 9$~Gyr ago, and the
metallicity had reached the solar value by 5 Gyr ago, consistent with
direct spectroscopy constraints from RGB stars
\citep[e.g.][]{guh06,kal06,gil09}.  In addition, all the GSS fields
probed thus far reveal a large spread in metallicity ($\geq 1.5$ dex).
Taken together, these properties suggest an early-type progenitor,
such as a dwarf elliptical galaxy or spiral bulge.  This further
supports inferences from $N$-body modelling which suggest that the GSS
progenitor was a fairly massive object with some degree of rotational
support.  Indeed, \citet{far13} argue it was likely the fourth or
fifth most massive Local Group galaxy as recently as 1~Gyr ago.

\subsection {Other Inner Halo Substructure}
\label{sec:inner}

\begin{figure}[h]
\begin{center}
\includegraphics[width=\textwidth]{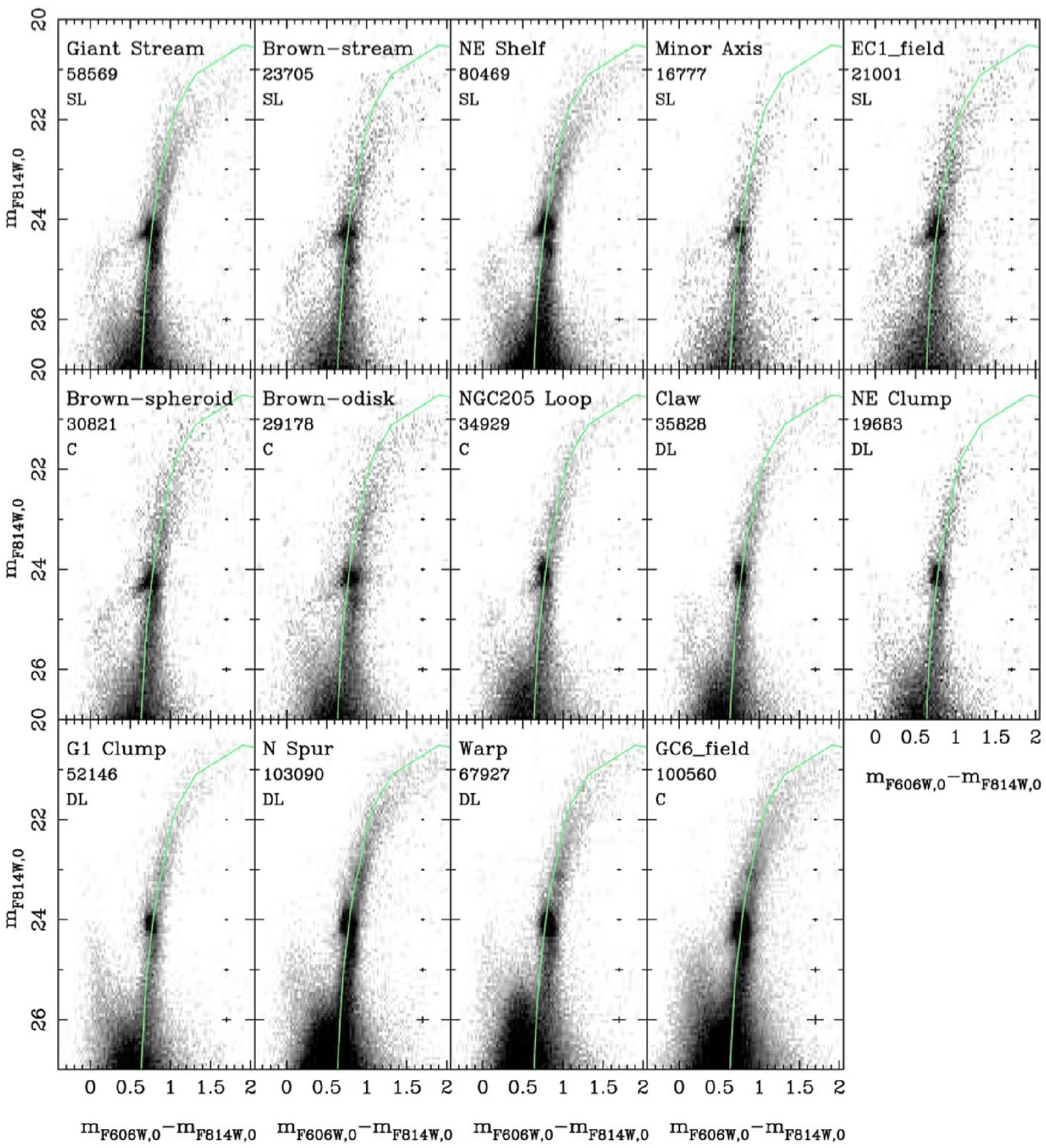}
\end{center}
\caption{HST/ACS Hess diagrams of 14 fields in the inner halo of M31,
  many of which lie on substructures seen in Fig. \ref{fig:inner_map}.
  The ridge line of 47 Tuc ([Fe/H]=-0.7 and age 12.5 Gyr) has been
  shifted to the distance of M31 and overlaid in each case.  Based on
  their CMD morphologies, the fields can be segregated into
  two classes; stream-like (SL) fields with tilted red clumps and
  extended horizontal branches and disk-like (DL) fields with round
  red clumps and a population of blue plume stars.  Composite (C)
  fields have features in common with both and likely represent cases
  where different material is projected along the line-of-sight.
  Reproduced from \citet{ric08}.}
\label{fig:jenny_cmds}
\end{figure}

While tidal debris from the GSS progenitor can explain the origin of
some of the M31 inner halo substructure, other features require a
different origin.  In particular, the substructures lying near the
major axis -- such as the NE and G1 Clumps and the Claw -- do not
arise naturally in models of the dissolution of the stream progenitor.

A first step to understanding this low latitude substructure was taken
by \citet{iba05} who analysed the kinematics of stars in numerous
fields around M31, including the G1 and NE Clumps. They noted a strong
signature of rotation in almost all fields out to a galactocentric
radius of $\sim40$~kpc, with some further detections out to $70$~kpc.
Stars are observed to move with velocities close to those expected for
circular orbits in the plane of the M31 disk and with a typical
velocity dispersion of 30~km~s$^{-1}$.  Based on stacked spectra,
\citet{iba05} estimated the mean metallicity of these rotating outer
populations to be [Fe/H]~$\sim-0.9$.

The irregular morphology yet coherent rotation observed in the outer
disk regions led \citet{iba05} to speculate that a vast disk-like
structure was being assembled as a result of multiple accretion
events.  However, this interpretation faced a number of challenges,
such as the homogeneity of the constituent stellar populations and the
need for the accreted satellites to be sufficiently massive so that
dynamical friction could circularise their orbits before disruption.

An alternative explanation is that much of the inner halo substructure
is material that has either been torn off of the M31 outer disk or
dynamically heated from the disk into the halo.  It has long been
known that the accretion of a low mass companion can have rather a
disruptive effect on a stellar disk \citep[e.g.][]{qui93,wal96} and
more recent work has quantified the way in which disk stars can get
ejected into the halo through such events \citep[e.g.][]{zol09,pur10}.
\citet{kaz08} demonstrated how the accretion of a population of
satellites with properties drawn from a cosmological simulation can
produce distinctive morphological features in the host galaxy's disk,
similar to the inner halo substructures seen in M31.  Additionally,
they confirmed earlier work that showed the final distribution of disk
stars exhibits a complex vertical structure that can be decomposed
into a thin and thick disk \cite[see also][]{vil08}.

\begin{figure}[h]
\begin{center}
\includegraphics[width=\textwidth]{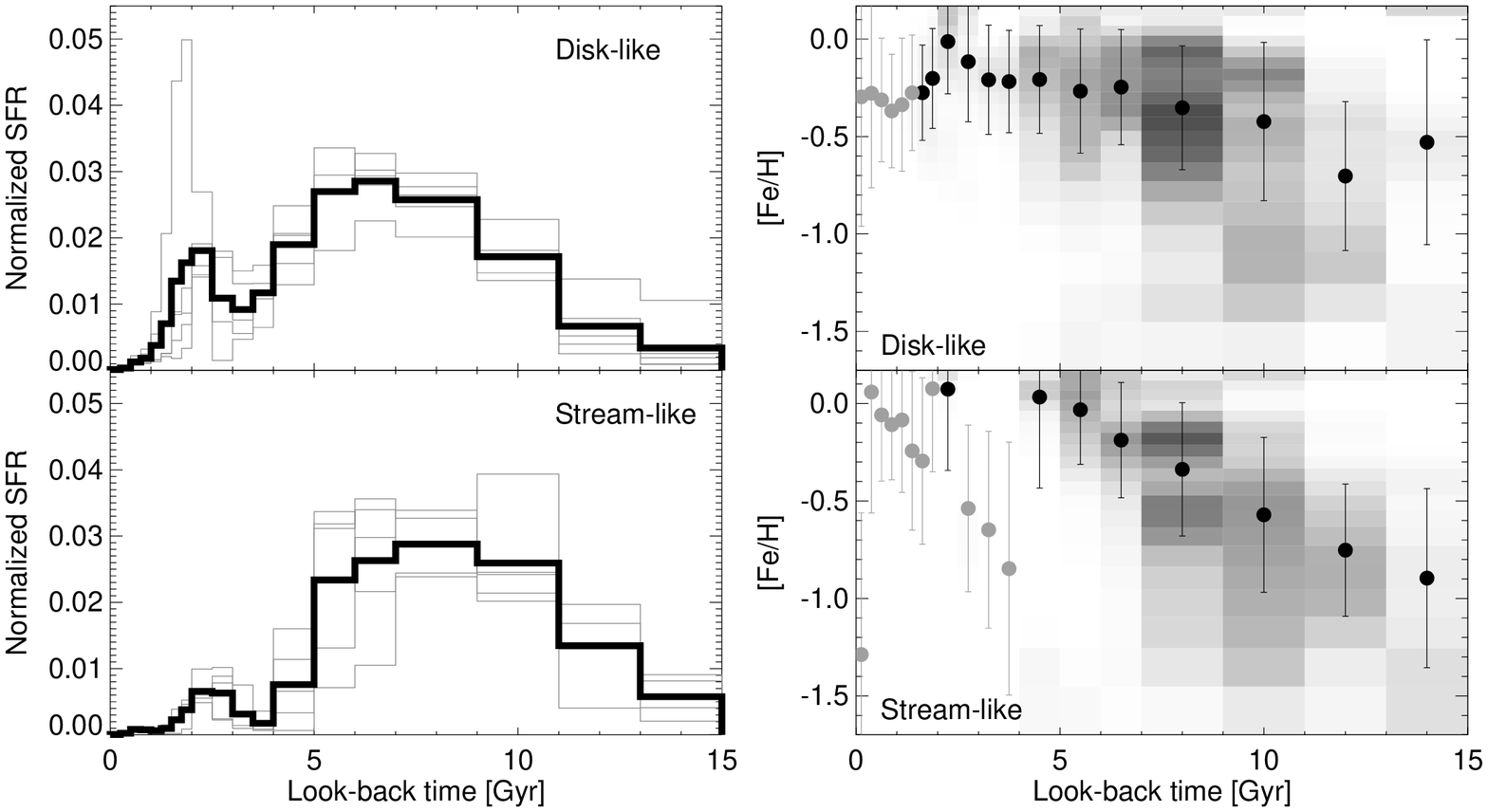}
\end{center}
\caption{(Left) The star formation histories (SFHs) and (Right)
  age-metallicity relations (AMRs) of M31 inner halo substructure
  fields, derived from quantitative fitting of the CMDs in Fig.
  \ref{fig:jenny_cmds}.  The SFHs for the individual fields are shown
  in light grey and are normalised to the total mass of stars formed
  in each field.  Overlaid in bold are the average behaviours of the
  normalised SFHs.  The filled circles in the AMRs show the median
  metallicity in each age bin, with grey circles indicating those bins
  which contain $\leq$1\% of the total stellar mass and hence carry
  significant uncertainties.  There is more star formation at early
  (late) times in the stream (disk) fields and the stream fields also
  exhibit a more rapid early chemical evolution. }
  \label{fig:sfhs}
\end{figure}

Examination of the constituent stellar populations in the non-GSS
debris fields supports the idea that this material has originated in
the disk \citep[e.g.][]{fer05,bro06,ric08,ber15a}.  The $`$DL' fields
in Fig.  \ref{fig:jenny_cmds} are rather homogeneous in appearance,
all displaying a round red clump with significant luminosity width, a
well-populated blue plume and no apparent horizontal branch --
features that indicate continuous star formation and a moderately
young mean age.  The quantitative SFHs and AMRs of the $`$disk-like'
debris fields further strengthen this assertion; roughly $\sim 65$\%
of the stars formed in the last 8~Gyr and chemical evolution proceeded
at a modest pace, starting from a pre-enriched level (see top panels of Fig.
\ref{fig:sfhs}). Most importantly, these trends are strikingly similar
to those that have been measured for populations in the M31 outer disk
 \citep{ber12,ber15b}.

It is also notable that both the stream-like and disk-like fields in
Fig. \ref{fig:sfhs} show evidence for an enhancement in the rate of
star formation roughly 2~Gyr ago. This is surprising given that
the constituent stellar populations in these fields have very
different origins, and that many of them are substantially displaced
from the main body (and the gas disk) of M31. There is now strong
evidence that the M31 outer disk underwent a burst of star formation
around this epoch, likely triggered by the relatively close passage of
M33 \citep{ber12,ber15b}. The existence of trace populations from this
episode scattered throughout the inner halo, including up to 20~kpc along the
minor axis, argues for a redistribution of disk material
in the intervening time.  It would thus appear that, in addition to
being heavily polluted by GSS debris, the M31 inner halo also contains
a widespread component of heated disk stars \citep{ber15a}. This
idea was independently raised by \citet{dor13} who find that a non-negligible fraction of the
inner halo stars identified kinematically in M31 show a luminosity
function consistent with an origin in the disk.  

Both the highly
structured nature of the outer disk {\it and} the presence of
displaced disk stars in the halo could be explained by one or more
violent accretion events.  Given the likely transitory nature of the outer disk
substructures \citep{iba05} and the fact that stars as young as 2~Gyr have
been displaced into the halo, the event responsible for disrupting
the disk must have been
rather a recent one and it is  tempting to speculate that it has been 
the head-on impact of the GSS progenitor roughly 1~Gyr ago
\citep[e.g.][]{mori08,sad14}.  If this scenario is correct, it implies
that, in spite of its extremely messy appearance, all of M31's inner
halo substructure can be traced to the direct and indirect effects of
a single event.

\subsection{Outer Halo Substructure}
\label{sec:outer}
The outer halo debris features in M31 are much more poorly understood
than those of the inner halo. While the inner halo is populated by
tidal debris with a variety of morphologies and moderate
metallicities, the outer halo is dominated by fairly narrow stellar
streams and arcs which are, with the exception of the metal-rich
component of Stream C, only apparent in maps constructed from
metal-poor stars (see Fig. \ref {fig:outer_map}).  The outer halo
streams are of such low surface brightness that detailed
characterisation of their stellar populations and kinematics has thus
far been difficult.  As will be discussed in the following
section, GCs offer a very exciting way to probe the tidal features in
these parts.
  
There are the four main structures visible in the far
outer halo -- Stream A, the E Cloud, the SW Cloud and the NW Stream.
All of these features lie at radii $\ga 100$~kpc from the center of
M31 and subtend at least a few tens of kpc in length.  Their
 CMDs  indicate a similar metallicity of
[Fe/H]~$\sim -1.3$ \citep{iba07,iba14,car11,bate14}.  The most
luminous of the outer halo structures is the SW Cloud which
\citet{bate14} estimate contains $\sim 5.6\times10^6$~L$_{\odot}$ or
equivalently $M_{V} \approx -12.1$.  This luminosity is approximately
75\% of that expected for the feature on the basis of its measured
metallicity and the \citet{kir11} luminosity-metallicity relation.
While this might indicate that a sizeable fraction of the luminosity
of the parent object has been detected, it remains unclear at present
whether these most distant halo features originate from distinct
accretion events or material torn off from a single progenitor.
Indeed, an interesting question is whether any of the outer halo
debris can be traced to the accretion of the GSS progenitor.  Although
the metallicity of these features is considerably lower than that of
the core of the GSS, it is a good match to that of the stream envelope
\citep{iba07,gil09}.

It is curious to note that the dwarf spheroidal satellite And
\rm{XXVII} appears projected on the upper segment of the NW
Stream.  Discovered by \citet{ric11}, this faint ($M_V \approx
-7.9$) system is highly morphologically disturbed and it is tempting
to speculate that it may be the source of the NW Stream debris.
However, the metallicity of the stream stars appears somewhat 
higher than that of the dwarf galaxy which complicates the
interpretation \citep{car11,col13}.  Furthermore, \citet{col13} note
additional kinematic substructure in the vicinity of And \rm{XXVII}
which is not yet understood.

\citet{iba14} have recently conducted a global analysis of the
large-scale structure of the M31 halo using data from the PAndAS
survey.  Despite the presence of copious substructures throughout,
they find that the stellar halo populations closely follow power-law
profiles that become steeper with increasing metallicity.  The {\it
  smooth} metal-poor halo component (defined as the population with
[Fe/H]~$ < -1.7$ that cannot be resolved into spatially distinct
substructures with PAndAS), has a global (3D) power-law slope of
$\gamma=-3.08\pm0.07$ and an almost spherical shape, but accounts for
a mere $\sim 5\%$ of the overall halo luminosity.  By far, most of the
luminosity of the halo out to the edge of the PAndAS survey resides in
moderate-metallicity substructure.  \citet{iba14} estimate that the
total stellar mass of the M31 halo at distances beyond $2^{\circ}$ is
$\sim 1.1\times 10^{10}$ M$_{\odot}$ and that the mean metallicity
decreases from [Fe/H]~$=-0.7$ at R~$= 30$~kpc to [Fe/H]~$ = -1.5$ at
R~$=150$~kpc for the full sample.  An alternative approach to studying
the outer halo in M31 has been taken by the SPLASH team who have
obtained spectroscopy of RGB stars in many pencil-beam fields
extending out to 175~kpc \citep{guh06,kal06,gil06,gil12,gil14}.  In
contrast to photometric studies, their approach allows them to
identify and remove kinematic substructure in their fields, at least
out to 90~kpc.  Although it is not possible to directly compare the
results of the PAndAS and SPLASH surveys, which are based on very
different sample selections, their inferences on the global halo
properties of M31 appear to be largely consistent \citep{iba14,gil14}.
 
\section{Globular Clusters as Probes of Tidal Streams in M31}
\label{sec:gcs}

An intriguing question regarding the substructures seen in
the halo of M31 is whether they show any degree of spatial correlation
with members of the M31 GC system.  This is motivated in part by the
long-held suspicion that a substantial number of the GCs in the Milky
Way halo did not form {\it in situ}, but rather in small satellite
dwarf galaxies that subsequently fell into the Galactic potential well
and disintegrated. This idea was first suggested in the seminal paper
by \citet{sea78} and was spectacularly verified in the early 1990s
with the discovery of the disrupting Sagittarius dwarf, which is in
the process of depositing {\it at least} five GCs into the outer halo
of the Milky Way \citep[e.g.][]{bel03,law10}. Modern studies of the
Galactic GC system have only served to add further weight to the
assertion -- it is now known that the abundances, velocities, ages,
horizontal branch morphologies, and sizes of perhaps up to a third of
Milky Way GCs are consistent with an external origin
\citep[e.g.,][]{zinn93,mac04,mac05,mar09,for10,dot10,dot11}.

Historically almost all work on the M31 GC system has been confined to
regions comparatively close to the galactic center, typically within
R~$\sim 20-25$\ kpc; however the situation has changed thanks to the
aforementioned wide-field mapping surveys of M31. In particular, the
INT/WFC and PAndAS surveys, but also to some extent the SDSS, have
facilitated the first detailed and uniform census of the outer halo GC
system of M31 \citep{hux05,hux08,hux11,hux14,mar06,zinn13,zinn14}.
Remote clusters are abundant in M31; there are now more than $90$
objects known to reside at projected radii beyond $25$\ kpc, $13$ of
which sit outside $100$\ kpc in projection, with the most the most
distant at $\sim 140$\ kpc in projection and up to $200$\ kpc in three
dimensions \citep{mac10a,mac13b}.  This is many more than is seen in
the outskirts of the Milky Way -- while the disparity in the number of
GCs in the Milky Way and M31 within $\sim 25$ kpc is roughly 3:1 in
favour of M31, outside this radius it is more like 7:1 in favour of
M31 \citep{hux14}.

\begin{figure}[h]
\begin{center}
\includegraphics[width=12cm]{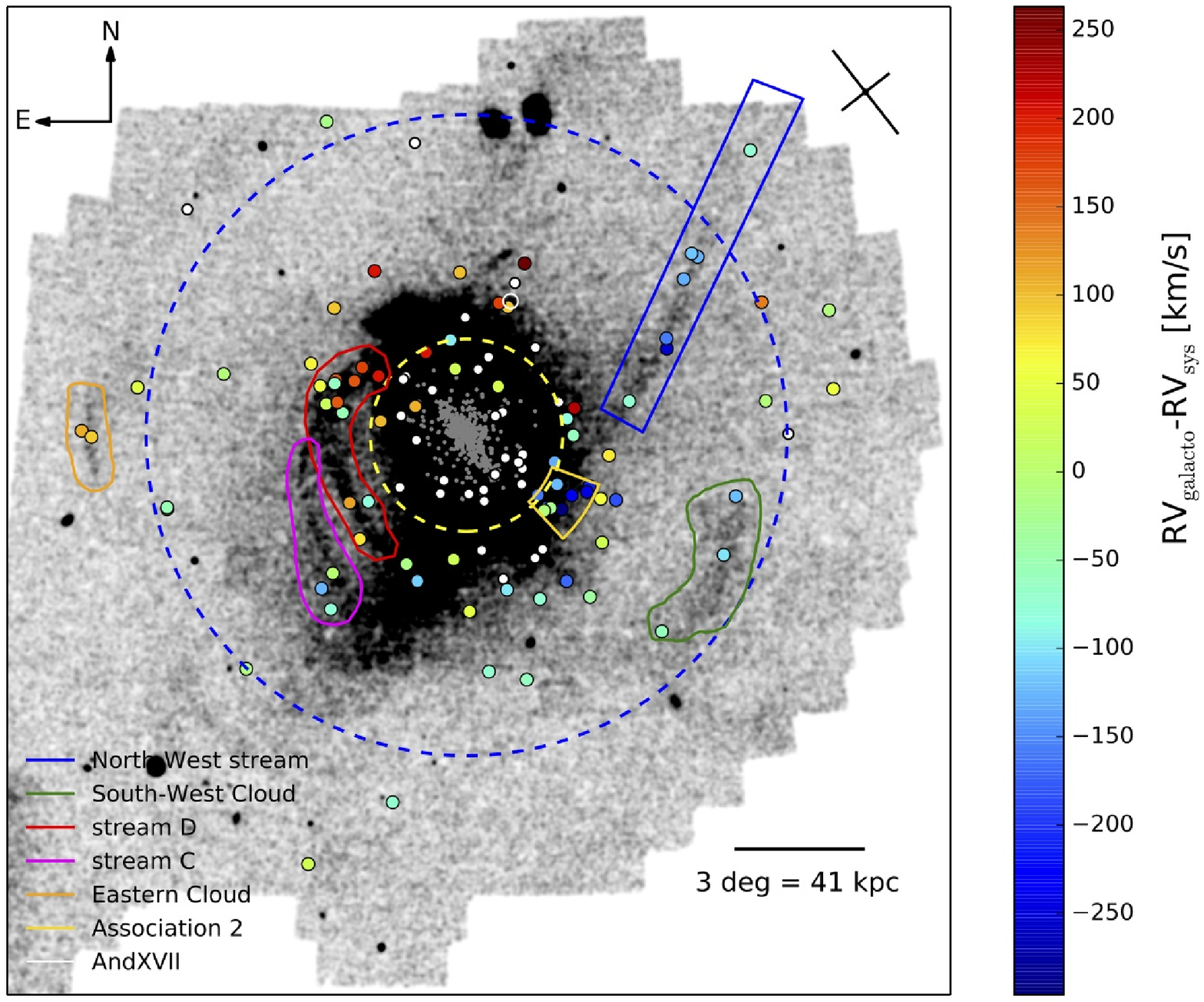}
\end{center}
\caption{Map showing extant radial velocity measurements for M31 outer
  halo GCs, projected on top of the PAndAS metal-poor RB map.  Most of
  the outer halo GCs can be seen to preferentially lie along stellar
  streams.  Inner halo GCs from the Revised Bologna Catalogue are
  shown as grey points. The GCs are colour-coded by their radial
  velocity in the M31-centric frame (white points indicate those
  objects with no radial velocity measurement), and the inner and
  outer dashed circles correspond to radii of 30 and 100~kpc.  A clear
  rotational signature is seen with GCs in the NE side of the galaxy
  receding while those in the SW quadrant approach us. Additionally,
  coherent velocities are seen for GCs which lie along specific debris
  features, strongly suggesting that the GCs have been brought into
  M31 along with their host galaxies.}
\label{fig:gcs}
\end{figure}

Figure {\ref{fig:gcs}} shows the positions of all known M31 GCs
plotted on top of the PAndAS spatial density map of metal-poor RGB
stars. In the outer parts of the halo, where large, coherent tidal
debris streams are readily distinguished, there is a striking
correlation between these features and the positions of a large
fraction ($\sim 50-80\%$) of the GCs. Substantial
numbers of clusters are seen projected onto the NW Stream, the SW
Cloud, the E Cloud, and the overlapping portion of Streams C and D.
There is, in addition, a statistically significant overdensity of
clusters (``Association 2'') sitting near the base of the NW Stream,
that cannot (as of yet) be identified with a visible tidal stream in
the field halo.

\citet{mac10b} have demonstrated that the probability of this global
alignment between clusters and streams arising randomly is low -- well
below $1\%$ for a GC system possessing an azimuthally uniform spatial
distribution. This implies that the observed coincidence represents a
genuine physical association and hence direct evidence that much of
the outer M31 GC system has been assembled via accretion.  Moreover,
at least some of the properties of the accreted M31 GCs appear to be
consistent with those exhibited by ostensibly accreted Galactic
members -- particular examples being those of younger ages
\citep{mac13a} and extended structures \citep{hux11,tanv12}.

The argument made by \citet{mac10b} is based entirely on statistical
grounds; to determine on an object-by-object basis which GCs are
associated (or not) with a given substructure requires kinematic
information.  The most extensive kinematic study of the M31 outer halo
GC system to date is by \citet{vel13,vel14}, who acquired spectra for
$71$ clusters outside a projected radius of $30$\ kpc (representing
86\% of the known population in these parts); the velocities of these
objects in the M31-centric frame are color-coded in Fig.
{\ref{fig:gcs}.  It can be readily seen that GCs projected onto a
  given outer halo substructure tend to exhibit correlated velocities
  (see left panel of Fig. \ref{fig:gckin} for an example). Those
  objects on the NW Stream and SW Cloud reveal strong velocity
  gradients from one end of the substructure to the other, while
  clusters on the E Cloud form a close group in phase space. Members
  of the Stream C/D overlap area, and those in Association 2 split
  into additional sub-groups by velocity.  A remarkable feature of
  many of the ensembles considered by \citet{vel14} is the coldness of
  their kinematics, with all GC groupings exhibiting velocity
  dispersions consistent with zero given the individual measurement
  uncertainties.  These results strongly reinforce the notion that a
  substantial fraction of the outer halo GC population of M31 has been
  accreted, and that these clusters trace the velocities of the tidal
  streams from their progenitor systems.  Indeed, while definitive
  measurements have thus far only been possible for two substructures,
  the velocities of the GCs sitting on Stream C and on the SW Cloud
  have been shown to be in excellent agreement with those of the
  underlying stream stars \citep{col09, mac14}.

  Fig. {\ref{fig:gcs} also demonstrates the surprising result that the
    M31 GC system as a whole possesses bulk rotation -- those GCs to
    the west of M31 appear to systematically possess negative
    velocities in the M31-centric frame, while those to the east
    typically have positive velocities.  \citet{vel14} compared a
    variety of kinematic models to the data and found a rotation
    amplitude of $86 \pm 17$\,km\,s$^{-1}$ around an axis aligned with
    the M31 optical minor axis provided the best match.  This rotation
    velocity is quite substantial -- for comparison, \citet{vel14}
    also found evidence that the velocity dispersion in the cluster
    system decreases from $129^{+22}_{-24}$\,km\,s$^{-1}$ at $30$\ kpc
    to roughly $75$\,km\,s$^{-1}$ at $100$\ kpc.  The right panel of
    Figure {\ref{fig:gckin}} further elucidates this rotation by
    showing the GC velocities in the M31-centric frame as a function
    of projected radius along the major axis. It is apparent that the
    rotation of the outer halo GC system is in the same sense as for
    the inner halo clusters (and indeed the M31 disk), albeit with
    smaller amplitude. Importantly, the rotation does not seem to be
    driven purely by clusters that sit on the particular halo
    substructures nor by those sitting well away from any underlying
    halo feature -- both groups apparently share in the pattern
    equally.

\begin{figure}[h]
\begin{center}
\subfloat{{\includegraphics[width=6cm]{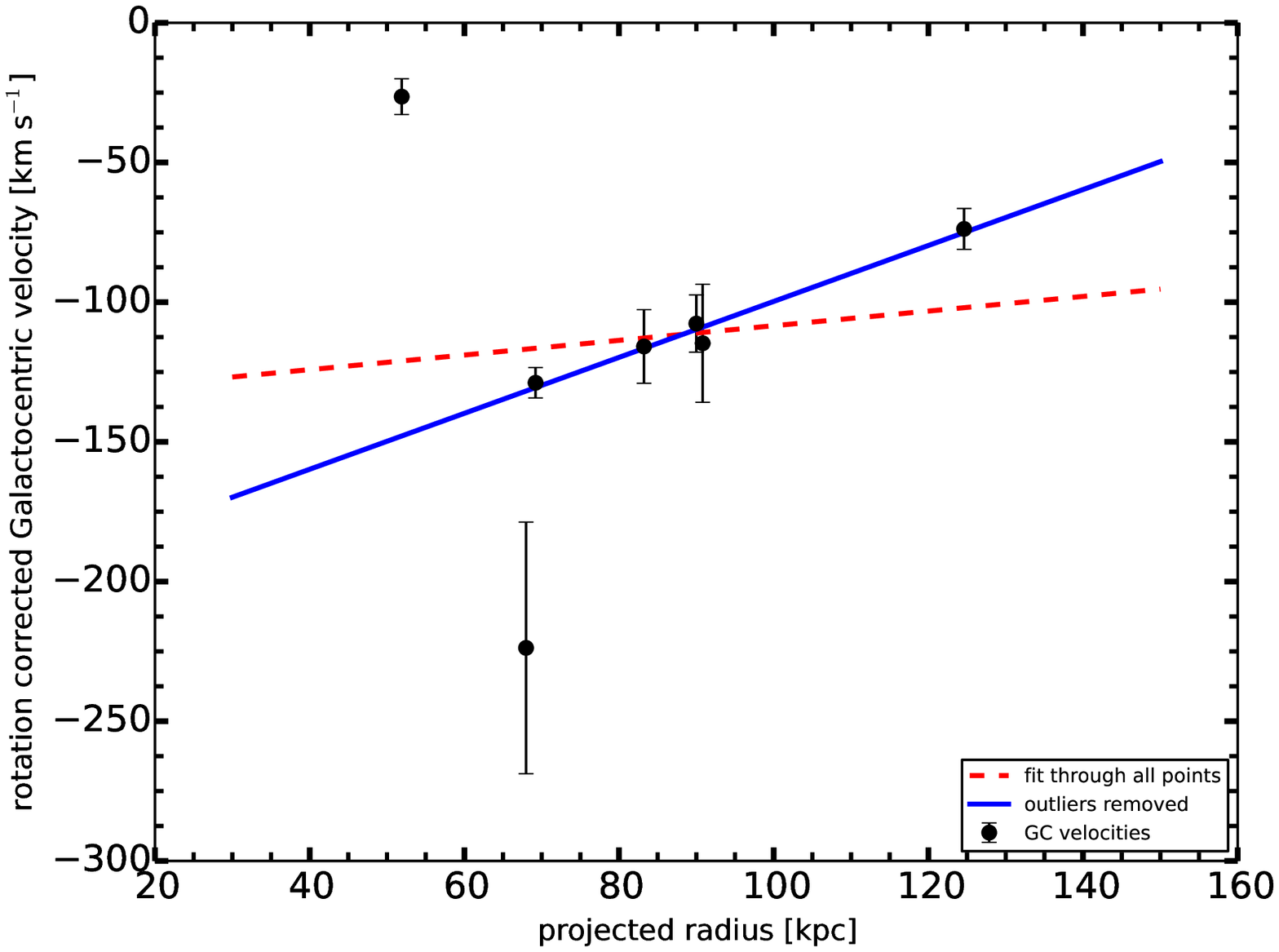}}}
\subfloat{{\includegraphics[width=6cm]{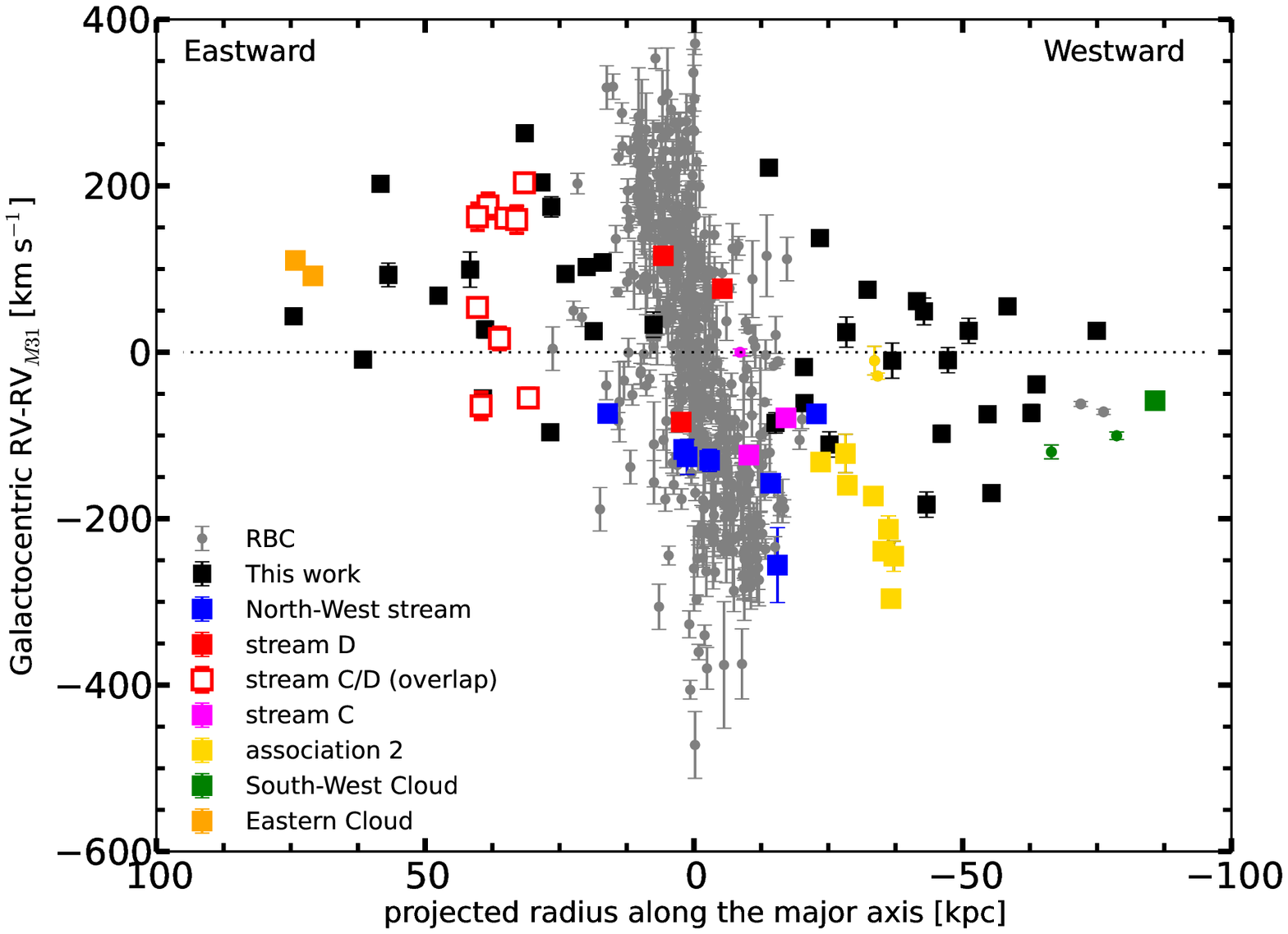}}}
\end{center}
\caption{(Left) Radial velocities for seven GCs that lie along the NW
  Stream.  A clear signature of radial infall is evident.  (Right) GC
  velocities, in the Galactocentric frame and corrected for the M31
  systemic motion, versus distance along the M31 major axis.  GCs
  which lie along specific debris features in the outer halo are
  colour-coded accordingly, while inner GC velocities taken from the
  literature are shown in light grey. The outer halo GCs rotate in the
  same sense as the inner GCs but with a somewhat smaller amplitude.
  The rotation is exhibited by the ensemble system of outer halo GCs
  and is not driven by specific stream features. Both panels are 
  reproduced from
  \citet{vel14}.  }
\label{fig:gckin}
\end{figure}

Understanding the origin of the angular momentum in the outer halo GC
system of M31 presents a significant challenge.  One possibility is if
a large fraction of the halo GC system was brought into M31 by just
one relatively massive host galaxy on a low inclination orbit.
However, in this scenario it is difficult to explain the observed
presence of distinct dynamically cold subgroups of GCs as well as the
typically narrow stellar debris streams in the halo. Another
possibility is that the outer halo GC system results from the
assimilation of several dwarf galaxies, but that these were accreted
onto M31 from a preferred direction on the sky.  This scenario might
well be related to the recent discovery that many dwarf galaxies, both
in the Milky Way and M31, appear to lie in thin rotating planar
configurations such that their angular momenta are correlated
\citep{iba13,paw12}. In this context it is relevant that almost all of
the dwarf galaxies thought to be members of the planes presently
observed in both M31 and the Milky Way are insufficiently massive to
host GCs, and furthermore that the rotation axes of the M31 outer halo
GC system and the M31 dwarf plane are misaligned with each other by
$\sim 45^{\circ}$.

\section{Tidal Streams from M31's Satellite Galaxies}
\label{sec:sats}

M31 has a substantially richer satellite system than the Milky Way.
Not only does it possess more dwarf companions, but it also possesses
several moderately luminous ones: M33 with $M_{\rm V}=-18.8$ and
the four dwarf ellipticals (dEs) M32, NGC~205, NGC~147 and
NGC~185 with $M_{\rm V}=-15.5$ to $-16.5$ \citep{mcc12,crn14}.
Tidal features are now known around nearly all of
these systems, indicating that they are in the process of
depositing tidally-stripped stars into the M31 halo.

The innermost satellites of M31 are M32 and NGC~205, which lie at
M31-centric distances of 25~kpc and 42~kpc, respectively
\citep{mcc12}.  These systems overlap the main body of M31 in
projection, a fact that has greatly complicated detailed analyses of
their outer structures.  Nonetheless, diffuse light studies indicate
that both systems exhibit breaks in their surface brightness profiles
which are accompanied by sharp changes in isophotal ellipticity and
position angle; such behaviour is consistent with
expectations for systems undergoing tidal interaction and stripping
\citep{choi02, joh02}.  Additionally, \citet{fer02} present isopleth
maps from a scanned deep photographic plate which reveal the
characteristic $``$S"-shape of tidal distortion in their peripheral
regions.  It is therefore curious that kinematic studies support
the existence of unbound stars at large radii in NGC~205 but not in
M32 \citep{geha06, how13}.

Using resolved star count data from the INT/WFC survey, \citet{mcc04}
 detected a 15 kpc long stellar arc that emanates
from the northern side of NGC~205 before bending eastward back to the
M31 disk.  This discovery resulted from the fact that the stellar loop
consists of RGB stars that are slightly bluer than the M31 inner halo
stars and hence have enhanced contrast on metal-poor star count maps.
Unfortunately, the true nature of this feature remains controversial.
While \citet{mcc04} detected a kinematic feature centered at
$-160$\,km\,s$^{-1}$ with a dispersion of $10$\,km\,s$^{-1}$ that they
attributed to the NGC~205 loop, subsequent re-examination of the data
by \citet{iba05} did not recover this cold component.  Instead, they
argued that the kinematic properties of this feature were consistent
with the bulk motion of M31's disk.  Furthermore, \citet{ric08} and
\citet{ber15a} find that deep HST CMDs of this feature can be
explained naturally by a combination of heated disk and GSS debris,
without requiring any additional component of stars, while
\citet{how08} find the most likely orbit of NGC~205 to be incompatible
with the location of the arc.

More recently, data from the the PAndAS survey has been used to
explore the outer regions of the dE satellites NGC~147 and NGC~185 to
extremely faint surface brightness levels using resolved stars.
Projected $\sim 100$~kpc north of M31, and lying at M31-centric
distances of 120~kpc and 180~kpc, respectively \citep{mcc12}, these
systems are sufficiently remote that contamination from M31 itself is
minimal and  the main complication is the presence of
substantial foreground Galactic populations along their sightlines.
\citet{crn14} trace both systems to $\mu_{g} \approx 32$
mag~arcsec$^{-2}$ and show that they have much greater extents than
previously recognized. As can be seen in Figs. \ref{fig:outer_map} and
\ref{fig:rod_halo}, NGC~185 retains a regular shape in its peripheral
regions while NGC~147 exhibits pronounced isophotal twisting due to
the emergence of long symmetric tidal tails. Even neglecting these
tails, NGC~147 appears more distended with an effective radius almost
three times that of NGC 185. In contrast to NGC~185, it also exhibits
no metallicity gradient. These differences in the structure and
stellar populations of the dEs suggest that tidal influences have
played an important role in governing the evolution of NGC 147, but
not NGC 185.  On
the assumption that NGC~147, NGC~185 and nearby dwarf spheroidal
Cass{\rm II} form a bound subgroup, \citet{ari16} show that it is
possible to find orbits around M31 which result in substantial tidal
disruption to NGC~147 but not the other two systems.

M31's most massive satellite is the low-mass spiral M33, lying
almost 210~kpc away \citep{mcc12}. Although no unusual structure was
detected around M33 in the early INT/WFC survey \citep{fer07}, the
deeper PAndAS data led to the discovery of a gigantic $``$S''-shaped
substructure that surrounds the main body of the galaxy
\citep{mcc09,mcc10}.  Traced to a projected radius of 40~kpc and
$\mu_{g} \approx 33$ mag~arcsec$^{-2}$, this feature coincides with a
similar feature that was detected in H$\,${\sc i} \citep{put09}.
\citet{mcc09} used $N$-body simulations to conduct a preliminary
exploration of the idea that this is the signature of M33's tidal
disruption as it orbits around M31.  They found that a relatively
close recent encounter could explain the appearance of this low
surface brightness substructure while satisfying the known phase
space constraints of the two systems.  Specifically, they suggest that
M31 and M33 came within 40--50~kpc of each other roughly 2.5~Gyr ago,
a hypothesis which is also consistent with the more recent Local Group
orbit modelling work of \citet{vdm12}.  This timescale is particularly
interesting since the outer disks of both systems appear to have
experienced strong bursts of star formation at this time and there is
evidence to suggest the inflow of metal-poor gas \citep{ber12,ber15b}.
Thus, while more detailed modelling of this interaction is
clearly required, there is tantalising evidence to suggest that it
could explain a number of puzzling aspects about the M31-M33 system,
such as the strong warps in both galaxies and the unusual synchronous
burst of star formation.

\section{Summary and Future Prospects}

The vast amount of work on the M31 halo over the past 15 years has led
to an exceptional situation where, in many ways, our knowledge of the
peripheral regions of that system far exceeds our knowledge of those
regions in our own Milky Way.  Studies have been able to identify and
characterise coherent debris streams in the M31 halo out to projected
galactocentric distances of $\sim 120$~kpc, and detect and map the
properties of the $`$smooth' halo out to $\sim 150-200$~kpc.  Even in
the inner halo ($\la 50$~kpc), we have a much clearer understanding of
the nature and origin of the tidal debris, thanks to our external
perspective. This can be contrasted with the situation in the Milky
Way where many of the inner halo/outer disk structures are very poorly
understood (e.g. the Monoceros, Hercules-Aquila, Tri-And, and Virgo
overdensities; see Chapters 3 \& 4).

In comparison to the Milky Way, the M31 inner halo is littered with
metal-rich debris. Some of this material has been stripped from the
GSS progenitor, while the rest is likely to be material from the M31
disk heated by the recent impact of this satellite.  M31 may also have
more outer halo tidal streams than the Milky Way, although most Milky
Way surveys to date have not had the sensitivity to detect
substructures in these parts.  Additionally, M31 (1) has a higher
fraction of its total  light in the halo component compared to
the Milky Way \citep[e.g.][]{iba14, gil12}}, (2) is characterised by a
smooth density profile unlike the broken power-law of the Milky Way
\citep{dea13} and (3) has a substantially larger population of halo
GCs, many of which lie along streams \citep{mac10b,vel14}.  It is
 likely that these differences result from the unique
accretion histories experienced by the two systems. M31 may have
experienced more accretions than the Milky Way, or it may simply have
experienced a more prolonged history of accretion.  Indeed, there is a
tantalising similarity between the properties of the Sagittarius dwarf
and those inferred for the GSS progenitor -- both are early-type
galaxies with estimated initial masses in the range $0.5-1 \times
10^9$~M$_{\odot}$.  The major building blocks of these halos
may well have been comparable, but their orbits and accretion times  
rather different.  In this spirit, it is interesting to speculate how the Milky
Way and M31 halos would have compared to each other $\sim 2$~Gyr ago,
before the GSS progenitor entered M31's inner halo.

A number of outstanding questions remain regarding the M31 system, and
much exciting progress can be expected over the next decade.  Some of
the most pressing issues that remain to be addressed include:

\begin{itemize}
\item{The development of a definitive understanding of M31's most
    significant recent accretion event: what was the true nature of
    the GSS progenitor and where is the remnant now?  How did it come
    to be on such an extreme orbit and how did it survive in this
    state until $\sim 1$~Gyr ago? What is the explanation for the
    second velocity component in the GSS and is there any connection
    between the GSS progenitor and the similarly metal-rich component
    of Stream C?}
\item{Understanding the origin of the outer halo debris. What and
    where are the progenitors of the outer halo debris streams? Do
    these features result from one or many accretion events? The
    associated GCs are likely to play an important role in answering
    these questions, and also in using these streams to derive refined
    estimates of the mass and potential of M31.  While line-of-sight
    distances to stream GCs are possible with HST observations, proper
    motion measurements may be possible for the most compact objects
    with Gaia.}
\item{Understanding the origin of the coherent rotation in the outer
    GC population, and how it can be reconciled with the supposedly
    chaotic accretion of parent dwarf galaxies into the halo.
    Additionally, do the underlying debris streams and the smooth
    field halo also exhibit this rotation?  Given the sparse nature of
    the stellar populations in these parts, very large-scale
    spectroscopy campaigns covering a significant fraction of the halo
    will be required to answer these questions. }
\item{Developing a holistic picture of M31's evolution that links its
    accretion history to its global evolution and current structure. There 
    are fascinating hints that the recent
    interaction and accretion history of M31 can explain a variety of
    puzzling observations. The close passage of M33 could excite the
    strong asymmetric warps and bursts of star formation observed in
    both systems, while the accretion of the GSS progenitor could
    further disrupt the M31 outer disk, displace some fraction of the
    disk stars into the halo and deposit a substantial amount of
    metal rich debris in the inner halo.  Much work is required to
    verify and fine-tune these ideas and there is a particular need
    for further detailed $N$-body modelling.  }
\item{Searching for the edge of the M31 stellar halo.  Tidal streams,
    field stars and GCs have been found to the very edge of the PAndAS
    survey, suggesting that they could extend yet further.  New
    wide-field mapping facilities such as Hyper-Suprime Cam on the
    Subaru telescope will be required to efficiently explore beyond
    the limit of PAndAS and may benefit from the use of specialised
    filters to discriminate between foreground and M31 populations.}
\end{itemize}

\acknowledgement
We thank Edouard Bernard and Jovan Veljanoski for their help in
creating Figs. \ref{fig:sfhs} and {\ref{fig:gcs}.   AMNF acknowledges
support from an STFC Consolidated Grant (ST/J001422/1) and
the hospitality of the Instituto de Astrofisica de Canarias while completing
this chapter. ADM is grateful for support from the Australian Research
Council through Discovery Projects DP1093431, DP120101237 and 
DP150103294.

\end{document}